\documentclass[12pt]{iopart}
\usepackage{graphicx}

\begin{document}

\title[Electric-field switchable magnetization via the
  Dzyaloshinskii-Moriya interaction]{Electric-field switchable
  magnetization via the Dzyaloshinskii-Moriya interaction: FeTiO$_3$
  versus BiFeO$_3$}

\author{Claude Ederer$^1$ and Craig J Fennie$^2$}

\address{$^1$ School of Physics, Trinity College, Dublin 2, Ireland}

\address{$^2$ Center for Nanoscale Materials, Argonne National
  Laboratory, Argonne, IL, USA}

\eads{\mailto{edererc@tcd.ie}, \mailto{fennie@anl.gov}}

\begin{abstract}
In this article we review and discuss a mechanism for coupling between
electric polarization and magnetization that can ultimately lead to
electric-field switchable magnetization. The basic idea is that a
ferroelectric distortion in an antiferromagnetic material can ``switch
on'' the Dzyaloshinskii-Moriya interaction which leads to a canting of
the antiferromagnetic sublattice magnetizations, and thus to a net
magnetization. This magnetization $\vec{M}$ is coupled to the
polarization $\vec{P}$ via a trilinear free energy contribution of the
form $\vec{P} \cdot (\vec{M} \times \vec{L})$, where $\vec{L}$ is the
antiferromagnetic order parameter. In particular, we discuss why such
an invariant is present in $R3c$ FeTiO$_3$ but not in the
isostructural multiferroic BiFeO$_3$. Finally, we construct symmetry
groups that in general allow for this kind of
ferroelectrically-induced weak ferromagnetism.
\end{abstract}

\section{Introduction}

One of the key challenges in the field of multiferroics is to design
and/or discover materials that exhibit a strong coupling between
magnetic and ferroelectric order parameters. Such materials are of
fundamental interest as they provide a novel platform to study how
microscopic degrees of freedom, such as spin and lattice, interact to
produce macroscopic phenomena.
These strongly coupled multiferroics are also anticipated to find
application in future generations of novel devices in which
magnetization can be controlled via an electric-field and/or electric
polarization can be controlled via a magnetic field.

Two different ways of controlling the state of a magneto-electric
system are possible: \emph{phase control}
\cite{Kimura_et_al_Nature:2003,Hur_et_al:2004,Lottermoser_et_al:2004}
or \emph{domain control} \cite{Fiebig_et_al:2002,Zhao_et_al:2006}. In
the first case an external field is used to trigger a phase transition
between two fundamentally different phases.
By tuning to the vicinity of a phase transition where two such phases
compete \cite{Newnham:1998, Tokura:2006:CMR}, e.g., an
antiferromagnetic-paraelectric phase and a ferromagnetic-ferroelectric
phase, a large magneto-electric response can be produced
\cite{Tokura:2006} even in systems where the intrinsic coupling
between magnetic and ferroelectric order parameters is not very strong
\cite{Fennie/Rabe_2:2006}.
In the case of domain control, the external field triggers a
transition between two equivalent, but macroscopically distinguishable
domain states, i.e., different realizations of the same phase. Here,
the magneto-electric coupling has to be large enough to overcome the
energy barrier for domain switching, which in general depends on both
the initial and the final domain. Once the field is removed, the
system is stable in the new domain state.

Several materials have been identified that realize one of the two
scenarios described above. In particular, much recent research has
been focused on systems where magnetic order itself breaks spatial
inversion symmetry and electric polarization therefore appears as a
secondary order parameter
\cite{Kimura_et_al_Nature:2003,Hur_et_al:2004,Lawes_et_al:2005}.
Several different microscopic models have been proposed that lead to
such ``magnetically-induced ferroelectricity.'' In some cases this
effect is caused by the presence of spin-orbit interaction
\cite{Katsura/Nagaosa/Balatsky:2005, Sergienko/Dagotto:2006}, whereas
in other cases spin-orbit coupling is not required
\cite{Sergienko/Sen/Dagotto:2006, Picozzi_et_al:2007}. See other
contributions in this Focus Issue for a more detailed discussion of
this interesting topic.

In the present article we discuss a somewhat different, but in a
certain sense complementary, possibility to realize coupling between
magnetic and ferroelectric order parameters, which can then be used to
achieve domain control of the corresponding multiferroic system. The
basic idea is that a ferroelectric distortion in an
antiferromagnetically ordered material can cause a small magnetization
due to the Dzyaloshinskii-Moriya interaction. In this case of
``ferroelectrically-induced ferromagnetism''\cite{Fox/Scott:1977} the
polar distortion gives rise to both the electric polarization and the
magnetization, hence the two quantities are inherently coupled.
Here, in contrast to the case of magnetically-induced ferroelectricity
mentioned above, the magnetization is the secondary order parameter
that is coupled to the primary order parameter, the electric
polarization.

The concepts and ideas reviewed and discussed in this article are
based mostly on a series of publications by the present authors, and
recently have been used to identify a specific class of materials that
are predicted to exhibit this effect
\cite{Ederer/Spaldin:2005,Ederer/Spaldin:2006,Fennie:2007}. The
concept of ferroelectrically-induced ferromagnetism was first
suggested by Fox and Scott \cite{Fox/Scott:1977} based on macroscopic
symmetry properties for the magneto-electric fluoride BaMnF$_4$. Here
we discuss a specific microscopic mechanism leading to such
macroscopic behaviour and analyze the corresponding symmetry
requirements.

In the following we first review the basic idea behind the proposed
mechanism, then summarize our previous work on this topic. We choose
to illustrate the general concept by discussing one structure in
detail, the ten atom rhombohedrally distorted ABO$_3$ perovskite
structure, paraelectric space group $R\bar{3}c$, and the corresponding
ferroelectric subgroup $R3c$, although the established principles are
easily generalizable. Specifically we show why the proposed effect is
present in magnetic $A$ site $R3c$ perovskites such as FeTiO$_3$, but
not in $R3c$ BiFeO$_3$, where the magnetic cations are situated on the
perovskite $B$ sites.  Finally, we present a rather general discussion
of the various symmetry aspects that have to be taken into account in
order for a material to exhibit the desired behaviour.
Ultimately our goal is to highlight the unique and powerful approach
of combining effective microscopic models with symmetry arguments to
guide first principles calculations in the discovery of new phenomena
and the design of their material realizations.

\section{Weak ferromagnetism and electric polarization}
\label{WFM}

It was realized by Dzyaloshinskii in 1957 that the appearance of
``weak'' ferromagnetism in some antiferromagnetic materials such as
e.g., Fe$_2$O$_3$, and its absence in the isostructural system
Cr$_2$O$_3$ can be explained solely on grounds of symmetry
\cite{Dzyaloshinskii:1957}. The symmetry of a magnetically ordered
material depends on the underlying crystallographic structure, the
orientation of the magnetic moments relative to each other, and on the
orientation of the individual magnetic moments with respect to the
crystallographic axes
\cite{Opechowski/Guccione:1965,Bradley/Cracknell:Book}. Dzyaloshinskii
showed that an invariant in the free energy expansion of the form
\begin{equation}
E^{DML} = \vec{D} \cdot (\vec{M} \times \vec{L}),
\label{DML}
\end{equation}
where $\vec{D}$ is a materials-specific vector coefficient, $\vec{M}$
is the magnetization, and $\vec{L}$ is an antiferromagnetic order
parameter, results in the secondary order parameter $\vec{M}$
appearing at the antiferromagnetic ordering temperature. In other
words, if the symmetry of the purely antiferromagnetic state is such
that the appearance of a small magnetization does not lead to a
further symmetry lowering, then any microscopic mechanism which
favours a nonzero magnetization, even if it is rather weak, will lead
to $\vec{M} \neq 0$.

\begin{figure}[t]
\centering
\includegraphics*[width=0.5\textwidth]{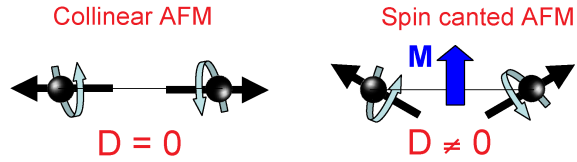}\\
\caption{\label{fig:canting} The presence of the Dzyaloshinskii-Moriya
  interaction ($\vec{D} \neq 0$) leads to a slight canting of the
  magnetic moments and a resulting net magnetization $\vec{M} \neq 0$
  in an otherwise collinear antiferromagnet.}
\end{figure}

It was later shown by Moriya that an invariant of the required form
can result from an antisymmetric microscopic coupling between two
localized magnetic moments $\vec{S}_i$ and $\vec{S}_j$:
\begin{equation}
E_{ij}^{\rm DM} = \vec{d}_{ij} \cdot \left( \vec{S}_i \times \vec{S}_{j}
\right) \quad ,
\label{DM}
\end{equation}
and that such an interaction arises from the interplay between
superexchange and spin-orbit coupling \cite{Moriya:1960}. Invariance
of the interaction energy (\ref{DM}) with respect to exchanging
$\vec{S}_i$ and $\vec{S}_j$ requires that $\vec{d}_{ij} = -
\vec{d}_{ji}$. The energy (\ref{DM}) is minimized when the two
magnetic moments form a 90$^\circ$ angle (or more accurately when
$\vec{d}_{ij}$, $\vec{S}_i$, and $\vec{S}_j$ form a left-handed system
for $|\vec{d}_{ij}| > 0$), but due to the simultaneous presence of the
generally much stronger Heisenberg-type interaction $E_{ij}^{\rm H} =
J_{ij} \vec{S}_i \cdot \vec{S_j}$, with $J_{ij} = J_{ji}$, which
favours either 0 or 180$^\circ$ angles, the presence of the
Dzyaloshinskii-Moriya (DM) interaction usually only leads to a small
canting between the interacting moments, i.e. a small deviation from
an overall collinear magnetic configuration as illustrated in
Fig.~\ref{fig:canting}.

Symmetry puts strict constraints on the DM vector $\vec{d}_{ij}$. In
particular, $\vec{d}_{ij}$ is identically zero if the midpoint between
the magnetic moments is an inversion center. Turning this around, it
means that in magnetic ferroelectrics, which do not exhibit any
inversion centers, the DM interaction can be expected to be a rather
common phenomenon (even though there can be other symmetry
restrictions which prohibit the DM interaction even in the absence of
any inversion centers). This raises the question of possible
cross-correlations between electric polarization and DM interaction in
ferroelectric magnets.

Many ferroelectric systems of interest today can be characterized by a
small polar structural distortion away from a centro-symmetric
reference structure. This reference structure is usually identical
with the crystal structure of the paraelectric phase but need not be
\cite{igartua}. The distortion, which leads to an electric dipole
moment, can be reversed by applying an appropriate electric field
(domain-switching). This leads to the following scenario
\cite{Ederer/Spaldin:2005,Fennie:2007}: If in the paraelectric
reference structure the midpoint between two neighbouring magnetic
ions is an inversion center, which is destroyed by the ferroelectric
distortion, then the ferroelectric distortion can ``switch on'' the DM
interaction between the two ions.\footnote{This has been called the
``structural-chemical criterion'' in Ref.~\cite{Fennie:2007}.}
In this case the materials specific parameter, $\vec{D}$, of
Dzyaloshinskii's invariant, Eq.~(\ref{DML}), can be identified with
the electric polarization $\vec{P}$ and the corresponding invariant in
the free energy expansion is:
\begin{equation}
E^{PML} \sim \vec{P} \cdot \left( \vec{M} \times \vec{L} \right) \quad .
\label{invariant}
\end{equation}
From Eq.~(\ref{invariant}) it can be seen that if it is possible to
reverse the direction of $\vec{P}$ (using an electric field) without
changing the orientation of $\vec{L}$, then the magnetization will
reverse too, in order to minimize the total free energy of the
system. Reversal of $\vec{L}$ can be prevented by sufficiently large
magnetic anisotropy, and thus an invariant of type (\ref{invariant})
opens up the possibility for electric field-induced magnetization
switching.

In the preceeding paragraph we have outlined the general scenario that
can lead to electric field-switchable weak ferromagnetism. In order to
find specific example materials that exhibit this effect it is
important to point out that even for cases where the ferroelectric
distortion destroys the inversion centers between adjacent magnetic
sites, and thus provides a necessary requirement for nonzero DM
interaction, there can be other symmetry operations that result in
$\vec{d}_{ij}=0$ or prevent the system from exhibiting a macroscopic
magnetization.\footnote{This has been termed the ``magnetic
criterion'' in Ref.~\cite{Fennie:2007}.} In the following we quickly
summarize our previous work along these lines and then discuss in
detail why the desired coupling is present in $R3c$ FeTiO$_3$ but
absent in BiFeO$_3$, even though these two systems have the same
crystallographic space group symmetry.

\section{First principles studies of BiFeO$_3$, BaNiF$_4$, and FeTiO$_3$}

The possibility of an electric-field switchable DM interaction was
first discussed in the context of BiFeO$_3$
\cite{Ederer/Spaldin:2005}. BiFeO$_3$ is an antiferromagnetic
ferroelectric with a N{\'e}el temperature of $\sim$ 643 K and a
ferroelectric Curie temperature of $\sim$ 1103 K
\cite{Kiselev/Ozerov/Zhdanov:1963,Teague/Gerson/James:1970}. It is
thus a very rare example of a multiferroic with both magnetic and
ferroelectric ordering temperatures above room temperature, and it is
probably the most-studied multiferroic to date. The primary magnetic
order in BiFeO$_3$ is ``G-type'' antiferromagnetism
\cite{Fischer_et_al:1980}, i.e., ``checkerboard''-like, but in
addition it has been reported that bulk single crystals exhibit a
superimposed cycloidal spiral magnetic ordering with a large period of
$\sim$ 620 \AA\
\cite{Sosnowska/Peterlin-Neumaier/Streichele:1982}. This spiral
ordering seems to be absent in thin film samples \cite{Bea_et_al:2007}
where instead a small magnetization has been found
\cite{Wang_et_al:2003,Eerenstein_et_al:2005,Bea_et_al:2007}. Early
reports of magnetizations up to 1$\mu_{\rm B}$/Fe
\cite{Wang_et_al:2003} could not be confirmed in other samples
\cite{Eerenstein_et_al:2005,Bea_et_al:2007} and therefore seem to be
caused by extrinsic effects.

In \cite{Ederer/Spaldin:2005} it was shown by first-principles
calculations that spatially homogeneous BiFeO$_3$ (without the spiral
spin structure) exhibits weak ferromagnetism as a result of the DM
interaction, and that the resulting magnetization is about 0.1
$\mu_{\rm B}$/Fe. Furthermore, it has been shown in
\cite{Ederer/Spaldin_2:2005} that the magnetization is only weakly
dependent on epitaxial strain, another indication that the very large
magnetization of about 1 $\mu_{\rm B}$/Fe found in the original thin
film samples \cite{Wang_et_al:2003} is most likely due to extrinsic
effects. The calculated magnetization agrees well with more recent
experimental thin film data
\cite{Eerenstein_et_al:2005,Bea_et_al:2007}. We point out that a
meaningful comparison between measured and calculated magnetization
requires that an antiferromagnetic mono-domain state has been achieved
in the experiment.

It was also shown in \cite{Ederer/Spaldin:2005} from explicit
first-principles calculations that the sign of the DM vector $\vec{D}$
in BiFeO$_3$ is independent of the polar distortion, but that it is
instead determined by a rotational (non-polar) distortion of the
oxygen octahedra network present in BiFeO$_3$. As we review in the
next section, this lack of coupling between the DM vector and the
polarization in BiFeO$_3$ is a question of symmetry
\cite{Ederer/Spaldin:2005,Fennie:2007} and not due to a ``weak
coupling'' as some authors have suggested
\cite{Sousa/Moore:2008,Sousa/Moore_2:2008}. Despite this absence of
coupling between $\vec{P}$ and $\vec{M}$ in BiFeO$_3$, it was
nevertheless realized that the DM vector can indeed couple linearly to
a structural distortion \cite{Ederer/Spaldin:2005}. However, the
relevant structural distortion in the case of BiFeO$_3$ (the
octahedral rotations) is non-polar, and therefore it was concluded
that electric-field switching of the weak magnetization in BiFeO$_3$
is unlikely as there is no obvious way to couple the electric-field to
the non-polar distortion. Still, it was suggested that if materials
can be found where the polarization and the DM vector are due to the
same structural distortion, then electric field-induced reversal of
the weak ferromagnetic moment is possible, and that there are no
general symmetry arguments that prevent such an effect. A specific
material that realized the predicted effect, however, remained elusive
until the work of Ref. \cite{Fennie:2007}.

A subsequent study of the antiferromagnetic ferroelectric BaNiF$_4$
revealed that the DM vector can indeed be proportional to a polar
ferroelectric distortion \cite{Ederer/Spaldin:2006}. However, the
overall symmetry of BaNiF$_4$ does not allow a macroscopic
magnetization, and it was shown that even though the DM interaction
leads to a local canting of neighbouring magnetic moments, all
components cancel out when taking the sum over the whole unit cell
such that $\vec{M}=0$. BaNiF$_4$ can therefore be classified as weak
\emph{anti}ferromagnet. It was pointed out that it should be possible
to switch the secondary antiferromagnetic order parameter in BaNiF$_4$
using an electric field \cite{Ederer/Spaldin:2006}.

Finally, it was recently shown in \cite{Fennie:2007}, that a series of
$R3c$ titanates, $A$TiO$_3$ with magnetic cations $A$=Mn, Fe, Ni, do
in fact combine all the symmetry properties necessary for
ferroelectrically-induced weak ferromagnetism. These compounds can be
synthesized at high pressure and remain metastable at ambient
conditions \cite{Ko/Prewitt:1988,Ming_et_al:2006}. It was confirmed in
\cite{Fennie:2007} by explicit first principles calculations that a
weak magnetization exists in these materials, and that it is reversed
when the polar distortion is reversed while keeping all other order
parameters (apart from the magnetization) fixed. It can therefore be
expected, that in these materials the magnetization can be reversed
using an electric field. This represents the first specific example
for the general mechanism of magneto-electric coupling outlined in
Sec.~\ref{WFM}

In \cite{Fennie:2007} two criteria for the rational design of
ferroelectrically-induced weak ferromagnetism were formulated: a
``structural-chemical criterion'' and a ``magnetic criterion''. As
already briefly remarked via footnotes in Sec.~\ref{WFM}, the
structural-chemical criterion implies that the midpoint between two
magnetic sites is an inversion center in the paraelectric reference
structure, whereas the magnetic criterion expresses the fact that
there should be no other symmetry elements besides these inversion
centers that prevent the system from exhibiting weak
ferromagnetism. In the following we will present a detailed comparison
between BiFeO$_3$ and FeTiO$_3$ (as representative for the titanate
systems discussed in \cite{Fennie:2007}), focusing on symmetry
aspects, and discuss why in FeTiO$_3$ both criteria are fulfilled,
i.e.  the DM interaction is induced by the ferroelectric distortion,
whereas in BiFeO$_3$ this is not the case. We will then develop
guidelines to construct symmetry groups that generally allow
ferroelectrically-induced weak ferromagnetism.

\section{ Magnetic $A$-site versus $B$-site $R3c$ distorted perovskites:
  symmetry aspects}
\label{sec:BFOvsFTO}

\begin{figure}[t]
 \centering
\includegraphics*[width=0.4\textwidth]{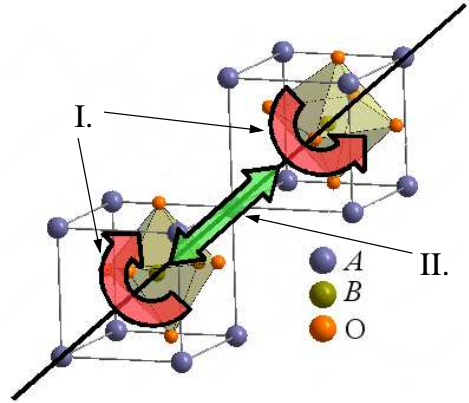}\\
\caption{\label{fig:R3c} Schematic depiction of the structural
  distortions that lead from the perfect cubic perovskite structure to
  the $R3c$ structure of BiFeO$_3$/FeTiO$_3$: I. counter-rotations of
  oxygen octahedra around [111] (red); II. ionic displacements along
  [111] (green)}
\end{figure}

The $R3c$ structure in which both BiFeO$_3$ and FeTiO$_3$ are found,
can be regarded as a distorted version of the 5-atom cubic perovskite
structure. The deviation from the ideal perovskite structure can be
decomposed into two components (see Fig.~\ref{fig:R3c}):
\begin{enumerate}
\item[I.]{antiferrodistortive counter-rotations of the oxygen
  octahedra around the [111] axis (leading to a unit cell doubling
  compared to the perovskite primitive unit cell), and}
\item[II.]{polar displacements of all the ionic sublattices relative
  to each other parallel to [111].}
\end{enumerate}
In FeTiO$_3$ the rotations (I) are so large that $R3c$ FeTiO$_3$ is
usually not considered to form a distorted perovskite structure but
rather the ``ferroelectric lithium niobate structure''.\footnote{We
point out that this distinction is not a question of symmetry but
merely of structural connectivity.} Note that this LiNbO$_3$ polymorph
of FeTiO$_3$ is structurally isomorphic to BiFeO$_3$ except that the
positions of the Fe and Ti/Bi cations are exchanged, i.e.:
FeTiO$_3$$\rightarrow$BiFeO$_3$ implies Fe$\rightarrow$Bi and
Ti$\rightarrow$Fe.

%
Let us first consider BiFeO$_3$ and FeTiO$_3$ in the ideal perovskite
structure, space group $Pm\bar{3}m$. Here we neglect both the
octahedral rotations (component I) and the ferroelectric displacements
(component II) while assuming that the spins order in a G-type
antiferromagnetic pattern. For both BiFeO$_3$ and FeTiO$_3$ this
paraelectric reference structure has inversion centers at all
midpoints between magnetic sites. Furthermore, these inversion centers
will be destroyed by any ferroelectric distortion, and thus the
structural criterion described in \cite{Fennie:2007} is fulfilled for
both systems.
However, since the cubic perovskite structure contains only a single
magnetic cation per unit cell, an additional symmetry operation exists
in the antiferromagnetically ordered state that requires $\vec{M}=0$,
i.e., the magnetic criterion is not fulfilled. The corresponding
symmetry operation translates all ions by one unit cell and then
inverts all magnetic moments through time inversion. In general,
$\vec{M}=0$ by symmetry, whenever the magnetic unit cell is a multiple
of the chemical unit cell. Specifically for the case of
BiFeO$_3$/FeTiO$_3$ this implies that the unit cell doubling caused by
the octahedral rotations (component I) is essential for obtaining weak
ferromagnetism and to fulfill the magnetic criterion. Therefore, the
octahedral rotations cannot be ignored in a proper symmetry analysis
of BiFeO$_3$ or FeTiO$_3$, and we conclude that cubic perovskite is
not a suitable paraelectric reference structure for obtaining
ferroelectrically-induced weak ferromagnetism.

In contrast, if we only neglect the polar displacements (component II)
while including the octahedral rotations (component I) in our
paraelectric reference structure, we obtain the Calcite or
``paraelectric lithium niobate'' structure (space group
$R\bar{3}c$). This is the closest centrosymmetric reference structure
for both BiFeO$_3$ and FeTiO$_3$ and it is thus the proper starting
point for a Landau free energy expansion describing the possible
coupling between $\vec{P}$, $\vec{L}$, and $\vec{M}$ in these systems
(see Sec.~\ref{sec:symmetry}).

\begin{figure}[t]
 \centering
\includegraphics*[width=0.5\textwidth]{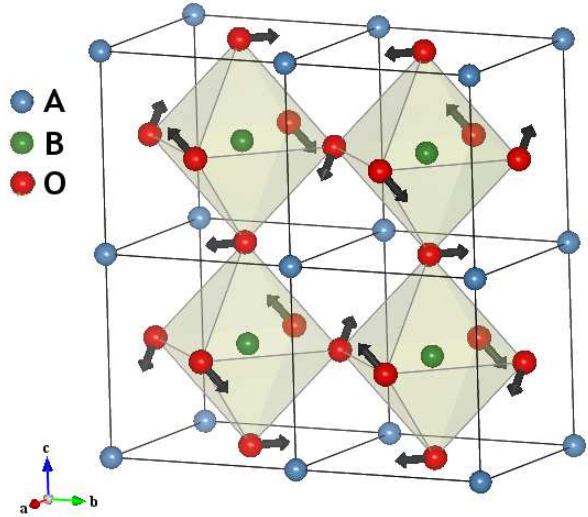}\\
\caption{Displacement vectors of the oxygen anions (red) corresponding
  to the octahedral rotations (I), relative to the ideal cubic
  perovskite structure. Due to the displacement of the oxygen anions
  located at the midpoints between adjacent $B$ sites (green), the
  corresponding inversion centers are destroyed, whereas the inversion
  centers located at the midpoints between the $A$ sites (blue) are
  conserved.}
\label{fig:rot}
\end{figure}

The crucial difference between BiFeO$_3$ and FeTiO$_3$ in the
$R\bar{3}c$ structure is the different local site symmetry of the
magnetic cations: in BiFeO$_3$ the magnetic cations (Fe$^{3+}$) are
situated on Wyckoff positions $2b$, corresponding to the $B$ site of
the underlying perovskite structure, whereas in FeTiO$_3$ the magnetic
cations (Fe$^{2+}$) occupy Wyckoff positions $2a$, corresponding to
the perovskite $A$ sites (see Fig.~\ref{fig:R3c}). To see whether the
structural criterion of \cite{Fennie:2007} is fulfilled, we have to
check whether the midpoints between the magnetic sites (i.e., the Fe
sites) in each system are inversion centers or not. As illustrated in
Fig.~\ref{fig:rot} this is the case for FeTiO$_3$ but not for
BiFeO$_3$, i.e., the structural criterion for
ferroelectrically-induced ferromagnetism is fulfilled in FeTiO$_3$ but
not in BiFeO$_3$.
Note that in the completely undistorted perovskite structure this
criterion was fulfilled for both systems, i.e., inversion centers were
located on the midpoints between the $A$ sites as well as on the
midpoints between the $B$ sites. The latter are destroyed by the
octahedral rotations (I). This is the reason why BiFeO$_3$ exhibits
weak ferromagnetism already in the paraelectric $R\bar{3}c$ phase,
i.e. weak ferromagnetism in BiFeO$_3$ is induced by the octahedral
rotations and not by the polar distortion \cite{Ederer/Spaldin:2005}.

We point out that in some previous publications ``$R3\bar{c}$''
symmetry has been used as basis for a free energy energy expansion of
BiFeO$_3$ (see e.g.
\cite{Vorobev_et_al_WFM:1995,Kadomtseva_et_al:2004,Sousa/Moore:2008}). ``$R3\bar{c}$''
is not standard space group notation\footnote{Note that here the
``bar'' indicating space inversion is combined with the $c$-type glide
plane and not with the threefold rotation as in $R\bar{3}c$}, and it
is therefore not fully clear what the corresponding symmetry
operations are, but apparently in this case it is assumed that
inversion centers are located \emph{between} the magnetic sites. As
stated above this is not true for BiFeO$_3$ in the proper paraelectric
reference structure. It follows that the symmetry analysis in
\cite{Vorobev_et_al_WFM:1995,Kadomtseva_et_al:2004,Sousa/Moore:2008}
in fact applies to FeTiO$_3$ instead of BiFeO$_3$.

It is now clear that only FeTiO$_3$, but not BiFeO$_3$, fulfills the
structural-chemical criterion. What remains to be discussed is whether
the inversion symmetry related to the midpoints between the magnetic
cations in FeTiO$_3$ is the only symmetry operation that requires
$\vec{M}=0$, i.e., whether the magnetic criterion is satisfied.  One
candidate for such a symmetry operation was already mentioned earlier:
the combination of a primitive translation of the paraelectric
structure with time inversion. However, this can only be a symmetry
operation if the magnetic unit cell is a multiple of the
crystallographic unit cell. As discussed above, the presence of the
octahedral rotations in $R\bar{3}c$ FeTiO$_3$ doubles the
crystallographic unit cell compared to simple perovskite, and thus
ensures that for the assumed G-type magnetic order the magnetic and
crystallographic unit cells are identical.

Apart from such ``magnetic Bravais lattice translations''
\cite{Bradley/Cracknell:Book} the fulfillment of the magnetic
criterion is mainly a question of how the individual magnetic moments
are oriented relative to the crystal axes, which is determined by the
magneto-crystalline anisotropy. As already shown by Dzyaloshinskii
\cite{Dzyaloshinskii:1957}, an orientation of the magnetic moments
perpendicular to the rhombohedral axis is required for weak
ferromagnetism to occur in crystallographic $R\bar{3}c$
symmetry. First principles calculations show that this orientation is
indeed favored by the magneto-crystalline anisotropy in FeTiO$_3$
\cite{Fennie:2007}, and thus the magnetic criterion is fulfilled for
this system.

In summary, we have shown that in FeTiO$_3$ both the
structural-chemical and the magnetic criteria for
ferroelectrically-induced weak feromagnetism are fulfilled. Due to the
different location of the magnetic cations in BiFeO$_3$ the structural
criterion is not fulfilled in this system (and thus the magnetic
criterion is not applicable). This shows that in FeTiO$_3$ the weak
magnetization is linearly coupled to the electric polarization whereas
this is not the case in BiFeO$_3$, as was also verified by explicit
first principles calculations in
\cite{Ederer/Spaldin:2005,Fennie:2007}.
It also becomes apparent that it is important to consider the full
crystallographic symmetry to analyze possible coupling between
$\vec{P}$ and $\vec{M}$, including also nonpolar structural
distortions such as the octahedral rotations in BiFeO$_3$/FeTiO$_3$.

\section{General symmetry considerations}
\label{sec:symmetry}

In the preceeding sections we described the general idea behind
ferroelectrically-induced weak ferromagnetism, we reviewed results for
some specific example materials, and presented a detailed comparison
of the two $R3c$ structure materials BiFeO$_3$ and FeTiO$_3$. In
Sec.~\ref{WFM} we mentioned that in order to achieve the desired
coupling between the polarization and the magnetization, the free
energy expansion of the high-symmetry phase has to contain a term of
the form (\ref{invariant}). In this final section we discuss some
general symmetry aspects and we list magnetic point groups that are
compatible with the existence of such a coupling between $\vec{P}$ and
$\vec{M}$.

According to the Landau theory of phase transitions, which describes
continuous transitions from a high symmetry configuration into a
configuration with lower symmetry, the free energy of the system can
be expanded in powers of the various order parameters
\cite{Landau/Lifshitz5:Book,Toledano/Toledano:Book}. Each individual
term in this expansion has to be invariant with respect to all the
symmetry operations of the high symmetry phase. Here, we are concerned
with a transition leading from a paramagnetic and nonpolar phase
(where $\vec{P}=\vec{M}=\vec{L}=0$) into a lower-symmetry phase where
all three order parameters are nonzero. This transition implicitly
also defines an intermediate nonpolar antiferromagnetic phase where
$\vec{L} \neq 0$ but $\vec{P}=\vec{M}=0$.

In order to achieve the desired coupling between $\vec{P}$ and
$\vec{M}$, the term $\vec{P} \cdot \left( \vec{M} \times \vec{L}
\right)$ has to be allowed in the free-energy expansion of the high
symmetry phase. Since the transformation properties of the polar
vector $\vec{P}$ and the axial vector $\vec{M}$ are well known, the
question of whether such a trilinear coupling between $\vec{P}$,
$\vec{M}$ and $\vec{L}$ is allowed, is basically a question about the
symmetry properties of the antiferromagnetic order parameter
$\vec{L}$. We point out that $\vec{L}$ cannot generally be classified
as either axial or polar vector; its symmetry properties depend on the
microscopic definition of $\vec{L}$ in terms of ionic magnetic moments
and the underlying crystallographic symmetry.  Ultimately, the design
criteria of \cite{Fennie:2007} aim at designing an antiferromagnetic
order parameter with the required macroscopic symmetry to couple
$\vec{P}$ and $\vec{M}$.

A first symmetry requirement for $\vec{L}$ has already been discussed
in Sec.~\ref{sec:BFOvsFTO}: the magnetic unit cell has to be identical
to the crystallographic unit cell. Otherwise, a symmetry element
exists which consists of time reversal combined with a lattice
translation of the paramagnetic phase. Such a transformation leaves
$\vec{L}$ and $\vec{P}$ invariant, but changes the sign of $\vec{M}$
and thus (\ref{invariant}) is not an invariant of the corresponding
high symmetry phase.  This symmetry requirement for $\vec{L}$ is
implicitly expressed in \cite{Fennie:2007} as part of the magnetic
criterion.

The next symmetry operation to consider is space inversion. The
presence of space inversion in the high symmetry group is not a
necessary requirement, but it simplifies the following analysis
considerably. We therefore restrict ourselves to cases where the
high-symmetry paramagnetic, non-polar reference structure is
centrosymmetric.
Notice, since $\vec{P}$ is a polar vector that changes sign under
space inversion and $\vec{M}$ is an axial vector that is invariant
under space inversion, that the antiferromagnetic vector has to be odd
under space inversion to allow an invariant of the form $\vec{P} \cdot
\left( \vec{M} \times \vec{L} \right)$. It is apparent that such an
antiferromagnetic order parameter arises for example if the inversion
centers of the high symmetry structure are located between two
antiferromagnetically coupled cations, but not if these inversion
centers are located on the magnetic cation sites themselves. This
symmetry requirement for $\vec{L}$ is thus related to the structural
criterion discussed previously.

We now construct symmetry groups that allow for an antiferromagnetic
order parameter with the two symmetry requirements outlined in the two
preceeding paragraphs. To simplify the presentation we only discuss
magnetic \emph{point} groups, not the full magnetic \emph{space} group
symmetry. Note that the macroscopic point group corresponding to a
particular microscopic space group is obtained by neglecting all
translational parts of the corresponding space group operations
\cite{Bradley/Cracknell:Book}. Therefore, the symmetry properties of
$\vec{L}$ outlined above are equivalent to stating that the magnetic
point group of the intermediate nonpolar antiferromagnetic phase can
contain neither space inversion nor time reversal symmetries
individually (i.e., neither $\bar{1}$ nor $1'$), but only the combined
operation of space inversion followed by time reversal (i.e.,
$\bar{1}'$). Note that these are also the symmetry requirements for
the existence of a magnetic toroidal moment $\vec{T}$ (see
\cite{Dubovik/Tugushev:1990,Ederer/Spaldin:2007}), which hints at a
close connection between the presence of a magnetic toroidal moment
and the presence of the magneto-electric coupling expressed in
(\ref{invariant}) (the exact nature of this connection will be the
subject of future investigations).

In addition, we realize that a free-energy invariant of the form
(\ref{invariant}) with an antiferromagnetic order parameter that can
be classified as a toroidal moment gives rise to an antisymmetric
linear magneto-electric effect (see e.g. \cite{Ederer/Spaldin:2007}),
and we can therefore now list all the magnetic space groups of the
desired intermediate state (where $\vec{L} \neq 0$ but
$\vec{P}=\vec{M}=0$). They are those which display an antisymmetric
linear magnetoelectric effect and contain the combined symmetry
operation of space inversion followed by time-reversal:
\begin{equation}
\label{sym1}
\bar{1}',\, 2/m',\, 2'/m,\, m'mm,\, 4/m',\,
\bar{3}',\, 6/m',\, 4/m'mm,\, \bar{3}'m,\, {\rm and}\ 6/m'mm \quad . 
\end{equation}
These are all possible space groups of the targeted antiferromagnetic
paraelectric phase, where a polar distortion can induce a weak
magnetization as a secondary order parameter. In order to construct
the possible polar subgroups we simply add a polar distortion along
one of the directions connecting antisymmetrical components of the
linear magnetoelectric tensor to all these groups, and calculate the
resulting symmetries. This results in the following 5 point groups:
\begin{equation}
\label{sym2}
1, \, 2',\, m,\, m', \, {\rm and}\ 2'm'm \quad .
\end{equation}
We note that these are precisely those determined by Fox and Scott by
considering all magneto-electric point groups which allow both a
spontaneous polarization and a spontaneous magnetization and requiring
$\vec{P} \perp \vec{M}$ \cite{Fox/Scott:1977}.

We point out that even though within our analysis we have first
proceeded from the paramagnetic-nonpolar case to the non-polar
antiferromagnetic case, and then subsequently to the polar-magnetic
case, it is not required that in the real system the magnetic phase
transition occurs at a higher temperature than the ferroelectric
transition. The important point is that an antiferromagnetic
paraelectric reference phase with the required symmetry can be
constructed in principle.
In fact displacive ferroelectric phase transitions often occur well
above room temperature, and therefore the critical temperature for the
effect described in this article is expected to be determined by the
antiferromagnetic N{\'e}el temperature in most cases.\footnote{In this
case, the magnetization will appear at $T_N$, rather than at the
ferroelectric transition, but the effect is still
ferroelectrically-induced.} The N{\'e}el temperature of
antiferromagnetic oxides with strong superexchange interaction is
often above room temperature and thus the described effect is not
limited to low temperatures.

\section{Summary and Outlook}

In this article we have illustrated the general principle behind
ferroelectrically-induced ferromagnetism through the DM interaction,
focusing on the corresponding symmetry requirements. We have discussed
in detail why these requirements are fulfilled for $R3c$ FeTiO$_3$,
where the magnetic Fe$^{2+}$ cations occupy the perovskite $A$ sites,
but not for isostructural BiFeO$_3$, where the magnetic Fe$^{3+}$
cations are located on the perovskite $B$ sites. As discussed in
Sec.~\ref{sec:BFOvsFTO}, the crucial difference is that in the
$R\bar{3}c$ paraelectric reference structure inversion centers are
located at the midpoints between the $A$ sites (structural criterion)
but not between the $B$ sites. These inversion centers are destroyed
by the ferroelectric displacements, and since they are the only
symmetry operations prohibiting a DM interaction between the $A$ sites
(magnetic criterion), the ferroelectric distortion ``switches on''
weak ferromagnetism in FeTiO$_3$. As a result, polarization and
magnetization in FeTiO$_3$ are coupled via the trilinear invariant,
Eq.~(\ref{invariant}).

In Sec.~\ref{sec:symmetry} we have reformulated the design criteria of
\cite{Fennie:2007} as symmetry requirements for the antiferromagnetic
order parameter $\vec{L}$, and we then constructed magnetic point
groups that are compatible with ferroelectrically-induced weak
ferromagnetism. In the most common case, where space inversion is a
symmetry element of the paraelectric reference structure, this
requires $\vec{L}$ to transform like a magnetic toroidal moment,
i.e. $\vec{L}$ has to be odd under both space and time inversion but
invariant under the combined operation.

Of course symmetry analysis gives only qualitative information, i.e.,
it tells us whether or not a certain effect is in principle possible
for a given symmetry. On the other hand, it is also desirable to
subsequently quantify the corresponding effect. Here, first principles
calculations using density functional theory represent an invaluable
tool. Within certain constraints, these calculations can provide very
reliable quantitative information about e.g., structural parameters,
lattice instabilities, magnetic coupling, and magnetization. First
principles calculations have been used in
\cite{Ederer/Spaldin:2005,Ederer/Spaldin:2006,Fennie:2007} to verify
and quantify weak magnetic order and its relation to structural
distortions, both polar and non-polar. Combined symmetry analysis and
first principles calculations have identified FeTiO$_3$ (and the
related Ni and Mn compounds) as prime candidate for the realization of
electric-field switchable weak ferromagnetism close to room
temperature \cite{Fennie:2007}, which now awaits experimental
verification.

Symmetry guided first principles design of novel promising materials
is in line with the general strategy for a rational computational
materials design outlined by Spaldin and Pickett
\cite{Spaldin/Pickett:2003}: after a candidate material with the right
symmetry has been found on symmetry grounds, first principles
calculations can be used to verify whether this material really shows
the desired effect and to determine the relevant quantities.

Such a strategy is particularly useful for the design of novel complex
oxide materials, where intrinsic properties can easily be hidden by
extrinsic effects, such as oxygen deficiency, micro-crystallinity, or
the presence of small amounts of competing phases, and can thus easily
be missed in a purely experimental approach. We therefore believe that
a rational materials design based on first principles calculations
indeed represents a very powerful approach to search for novel
materials with unexpected and technologically useful properties.

\ack C.E. acknowledges financial support by Science Foundation Ireland
and the Irish National Development Plan. Work at the Center for
Nanoscale Materials was supported by US DOE, Office of Science, Basic
Energy Sciences under Contract No. DE-AC02-06CH11357.

\section*{References}

\bibliographystyle{unsrt}
\bibliography{references.bib}

\end{document}